\documentclass{article}
\usepackage{graphicx} 
\usepackage{authblk}
\usepackage{hyperref}
\usepackage[top=1in, bottom=1.5in, left=0.75in, right=0.75in]{geometry}
\usepackage{natbib}
\usepackage{xcolor} 
\usepackage{amsmath}
\hypersetup{
    colorlinks=true,    
    linkcolor=blue,    
    citecolor=blue,    
    urlcolor=blue      
}

\title{Local and Remote Forcing Factors of Heatwave in India \\
-- A Reanalysis and Adjoint model based study}
\date{}

\author[1]{Abhirup Banerjee\thanks{Corresponding author: \href{mailto:abhirup.banerjee@uni-hamburg.de}{abhirup.banerjee@uni-hamburg.de}}}
\author[1]{Armin Koehl}
\author[2]{Frank Lunkeit}
\author[1]{Detlef Stammer}

\affil[1]{Institut f\"ur Meereskunde, Centrum für Erdsystemforschung und Nachhaltigkeit, Universit\"at Hamburg, Hamburg, Germany}
\affil[2]{Meteorologisches Institut, Centrum für Erdsystemforschung und Nachhaltigkeit, Universit\"at Hamburg, Hamburg, Germany}

\begin{document}

\maketitle

\begin{center}
\colorbox{gray!20}{
  \begin{minipage}{0.95\textwidth}
    \begin{center}
    {\large This work has been submitted to the Journal of Climate. Copyright in this work may be transferred without further notice.}
    \end{center}
  \end{minipage}
}
\end{center}

\vspace{4.5cm} 

\begin{abstract}
   Continental heatwaves can dramatically impact ecosystems and societies, e.g., by leading to excess mortality, wildfires, and harvest failures. With a warming climate, their impacts potentially intensify globally, but the Indian subcontinent appears to be particularly vulnerable to such extreme events. In this study, we use reanalysis and the adjoint of the atmospheric model, PlaSim, to identify drivers of heatwaves occurring April and May over north-central India. Reanalysis results suggest that the existence of high temperatures in the study region is highly sensitive to the low local soil moisture which is observed weeks before a heatwave commences. Soil moisture variability in northern India is influenced by moisture transport from the west during winter--spring. Preceding dry soil moisture conditions can be associated with a `persistent jet' conditions linked to atmospheric dynamical changes in the North Atlantic region. An associated northward shift in the upper tropospheric zonal wind occurs approximately a month prior to the heatwaves, influencing the area and intensity of western disturbances embedded in the jet stream. This weakens the moisture flow from the north of the Arabian Sea, further reducing soil moisture levels and creating conditions conducive to heatwaves. An adjoint sensitivity analysis and forward model perturbation experiments confirm the causal relationships for the proposed heatwave development mechanism over north-central India, identifying the remote influence of North Atlantic sea surface temperature variability on extreme temperatures in India. Our findings highlight the complex interplay of local and remote factors in heatwave development over India.
\end{abstract}

\section{Introduction}
Continental heatwaves (HWs) are meteorological phenomena characterized by prolonged periods of abnormally high temperatures that usually persist for days to weeks. These extreme events present a formidable hazard to ecosystems and societies, resulting in substantial economic losses and an alarming increase in mortality rates~\citep{zuo2015impacts, garcia2021current, campbell2018heatwave}. As an example, at the time of writing, the losses of life in India during 2024 already exceed 40,000 due to suspected heat stroke cases, and more than 100 heat-related deaths\footnote{\url{https://www.reuters.com/world/india/unrelenting-heatwave-kills-five-indian-capital-2024-06-19/}}. The influence of anthropogenic warming on the occurrence of HWs is evident from the increment in their occurrence and in their increased intensity  \citep{russo2014magnitude,domeisen2023prediction}. A significant upward trajectory in both the frequency and the severity of heat events is observed in the Indian subcontinent~\citep{panda2017increasing} posing significant risk for public health and societal well-being \citep{debnath2023lethal, im2017deadly, mazdiyasni2017increasing, de2024impact}. 

The development of HWs over India is a multifaceted process influenced by a range of factors, including anomalies in both regional and local atmospheric conditions~\citep{ratnam2016anatomy,dubey2021understanding}. In addition, large-scale teleconnections such as the El Niño-Southern Oscillation~\citep{hari2022strong,lekshmi2022modes,pai2022impact,kumar2023enhanced}, oceanic influences~\citep{pai2013long,vittal2020role}, and other regional factors~\citep{ganeshi2020understanding,satyanarayana2020phenology} also play a significant role in the occurrence of HWs across India. Specifically, \cite{vittal2020role} demonstrated that the Atlantic Ocean sea surface temperature (SST) had a greater influence on temperature extremes over India compared to the Pacific and Indian Ocean SST during the 1961--2010 period, with this influence being exacerbated by anthropogenic forcing. However, the detailed knowledge of the necessary conditions and the sequence of processes leading to HWs remains limited. Predicting HWs remains therefore difficult and relies primarily on statistical relations rather than the knowledge of physical processes.

Previous studies have shown that HWs are strongly associated with high-pressure systems and anticyclonic circulation, leading to moisture subsidence and clear skies that support heating \citep{barriopedro2023heat,perkins2015review,horton2016review,pfahl2012quantifying}. These summer temperature extremes can be amplified or dampened by land-atmosphere interactions. Numerous studies have shown that land-atmosphere interactions can amplify HWs \citep{lorenz2010persistence,jaeger2011impact,vogel2017regional} with soil moisture (SM) being a key mediator of these interactions \citep{seneviratne2010investigating,horton2015contribution}. Deficits in SM have been linked to specific extreme heat events also in the United States and Europe \citep{fischer2007soil,wehrli2019identifying}. In the context of temperature extremes in India, \citet{ganeshi2020understanding,ganeshi2023soil} have reported the role of SM variability in a warming climate. 

In recent years, the region of north-central India (NCI) has been particularly exposed to HWs and experiences a rising trend in the occurrence of HWs~\citep{rohini2016variability,rohini2019future}. During winter--spring, northwestern India receives moisture primarily from the north of the Arabian Sea. This is associated with `western disturbances' (WDs) which are cyclonic storms embedded in the subtropical westerly jet (SWJ)~\citep{yadav2012characteristic,hunt2018extreme,hunt2024increasing,baudouin2021synoptic}. The variability in moisture transport over northern India during this period is generally linked to  fluctuations in the intensity and frequency of WDs~\citep{hunt2024western}. More details about WDs can be found in ~\citet{hunt2018evolution}.

In this paper, we identify the potential contribution of local soil moisture conditions to the development of HWs during April and May over NCI, and the moisture dynamics in the preceding winter--spring which lead to such conditions. Specifically, we seek the answer to the question of what are the anomalous changes in the large-scale circulation patterns during winter--spring leading to drier SM which in turn creates conditions ideal for HW development during summer. We combine reanalysis-based investigations  with an adjoint sensitivity analysis via the adjoint of the atmospheric model PlaSim~\citep{fraedrich2005planet,blessing2014testing} which complement each other in identifying the distinctive precursory features of different climate variables prior to HWs and also  unravel the possible causal relationships among them. Such a holistic approach offers a comprehensive understanding of the physical mechanism underlying HW developments over NCI. Understanding the preconditions of HWs is crucial for their accurate prediction, preparation, and response, as such factors can either intensify or alleviate these extreme events. 

The remainder of the article is organized as follows. In Section \ref{data_method}, we introduce the methodology and data sets used. This includes the definition of HW events and the description of the model PlaSim and its adjoint used in this study. In Section \ref{results}, we present our findings from reanalysis data, followed by those from the adjoint-sensitivity study and lastly from the model forward run experiments provided in Section 4. Concluding remarks are provided in Section 5.

\section{Methodology}\label{data_method}

\begin{figure}[ht]
\centering
\includegraphics[width=33pc]
{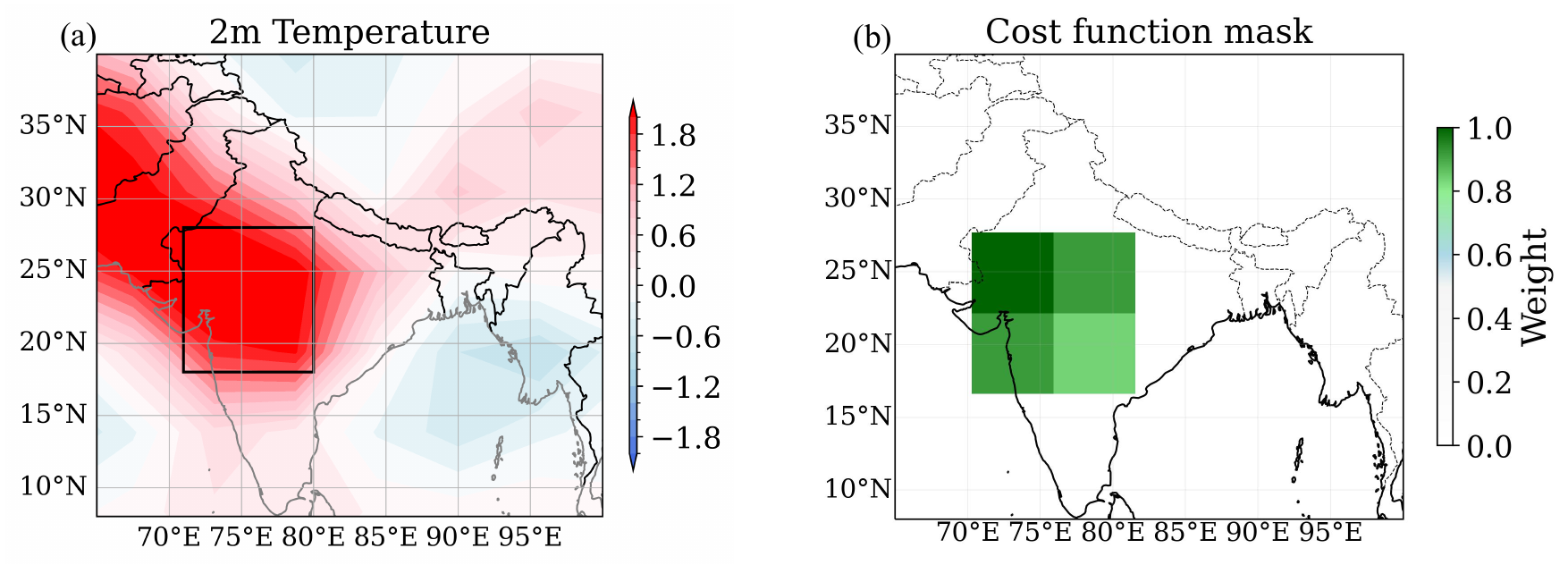}
\caption{(a) Temperature anomaly $[^\circ C]$ associated to the HWs 
over NCI in PlaSim. (b) Weighted mask used to compute the 
cost function.}
\label{fig:ad01}
\end{figure}

Several definitions of HWs exist in the literature~\citep{barriopedro2023heat}. This study is based on the HW indices ($HW_t$) over north-central India (NCI, $22^\circ-31^\circ$N, $70^\circ-77^\circ$E) provided by the India Meteorological Department (IMD)\footnote{\url{https://imdpune.gov.in/Reports/Met_Monograph_Cold_Heat_Waves.pdf}}. To compute the temperature anomaly associated with HWs over India, we use the $1^\circ\times1^\circ$ gridded maximum temperature (Tmax) daily data provided by IMD~\citep{srivastava2009development} which we accessed using IMDLIB python package~\citep{nandi2020iamsaswata}. To detect HW event in PlaSim, we considered when temperature (2m temperature, $T2m$) is above the $90^{th}$ percentile for at least 5 consecutive days. A typical example of a NCI HW pattern in PlaSim is shown in Figure~\ref{fig:ad01}a  representing the composites of $T2m$. 

To analyze the state of the atmosphere and its variability surrounding HW events between April and May over the period of 1980 to 2022, a detrended standardized composite anomaly and Empirical Orthogonal Function (EOF) analysis~\citep{dawson2016eofs} of different variables are used. For atmospheric variables, we use the NCEP-NCAR Reanalysis 1 data~\citep{kalnay2018ncep}. GLEAM v3.8a daily data is used for surface level soil moisture~\citep{martens2017gleam}. With respect to sea surface temperature data, the NOAA Extended Reconstructed sea surface temperature data (V5) data is used~\citep{huang2017noaa}. The dynamics of moisture transport associated with HWs will also be analyzed using the western disturbances (WDs) catalogue provided by~\cite{hunt2023linking} wherein the authors identified the regions with positive vorticity by analyzing the mean relative vorticity within 450-300 hPa and linked those regions based on a set of physical constraints.  


Adjoint sensitivities of Indian HWs will be analyzed to infer mechanisms that potentially can contribute to a HW development. Subsequently, our findings from both the reanalysis and adjoint sensitivity studies will be tested through forward perturbation experiments. For both steps, an intermediate complexity climate model PlaSim~\citep{fraedrich2005planet} is being used implemented globally with $5.6^\circ \times5.6^\circ$ horizontal resolution and 10 vertical levels.  The details and the parameters of the model can be found here\footnote{\url{https://www.mi.uni-hamburg.de/en/arbeitsgruppen/theoretische-meteorologie/modelle/plasim.html}}. 

The cost function is defined as the area weighted mean of daily mean 2m air temperature ($T2m$) for each of the HW onset (day 0) in PlaSim as depicted by the black rectangle in Figure~\ref{fig:ad01}a, 
\begin{equation}\label{eq1}
J = \frac{\int_A T2m W \, dA}{\int_A W \, dA}    
\end{equation}

Here $dA$ is the area of the black rectangle in the Figure \ref{fig:ad01}a; the weighting matrix $W$ is shown in Figure \ref{fig:ad01}b. 

The PlaSim adjoint is being generated using 
the automatic differentiation tool 
TAF~\citep{blessing2014testing,lyu2018adjoint}.
Adjoint models calculate the sensitivities of 
a cost function to various controls, such as 
initial and boundary conditions, revealing the 
non-local (in space and time) influence of 
model parameter on the specified cost 
function; examples of adjoint sensitivity 
analyses can be found 
by~\citep{marotzke1999construction,stammer2018pilot}. \cite{stammer2018pilot,kohl2019seasonal} used the adjoint of PlaSim coupled to the MITgcm (called the CESAM model) to study the sensitivity of temperatures over norther Europe to North Atlantic parameters.

The uncoupled PlaSim adjoint is being used 
here to find the sensitivity of extreme 
temperature shown in Figure~\ref{fig:ad01}a 
and defined in eq.(\ref{eq1}) with respect to 
other local and non-local atmospheric and land 
surface parameters, such as surface 
temperature, soil moisture or atmospheric 
transports. Since adjoint sensitivities are 
prone to developing fast-growing unstable 
modes limiting the length of the application 
period, Gaussian filters are used to dampen the 
fastest growing modes causing local spikes in 
the adjoint sensitivities and extend the 
window to some extent~\citep{stammer2018pilot}. To obtain adjoint 
sensitivities to the above cost function eq.(\ref{eq1}), the adjoint model is integrated backward 
in time.  Due to the exponentially growing 
modes in the backward (in time) running 
adjoint model, we restrict our sensitivity
analysis to 6 days backward from the
occurrence of a HW. 

Our analysis examines the sensitivity of extreme high temperatures with respect to the sea surface temperature (SST), upper tropospheric zonal wind ($Uwnd$), and soil moisture (SM). To this end we run the PlaSim model and its adjoint forward and backward for six distinct heatwave (HW) events, each corresponding to a different model year. We constructed the ensemble sensitivity by running the model backward using six independent initial conditions. The adjoint sensitivities presented below are expressed as ensemble means of these individual runs.

\section{Unveiling the NCI Heat Wave Dynamics: Reanalysis-based study}\label{results}
\begin{figure}[!ht]
\centering
\includegraphics[scale=0.6]{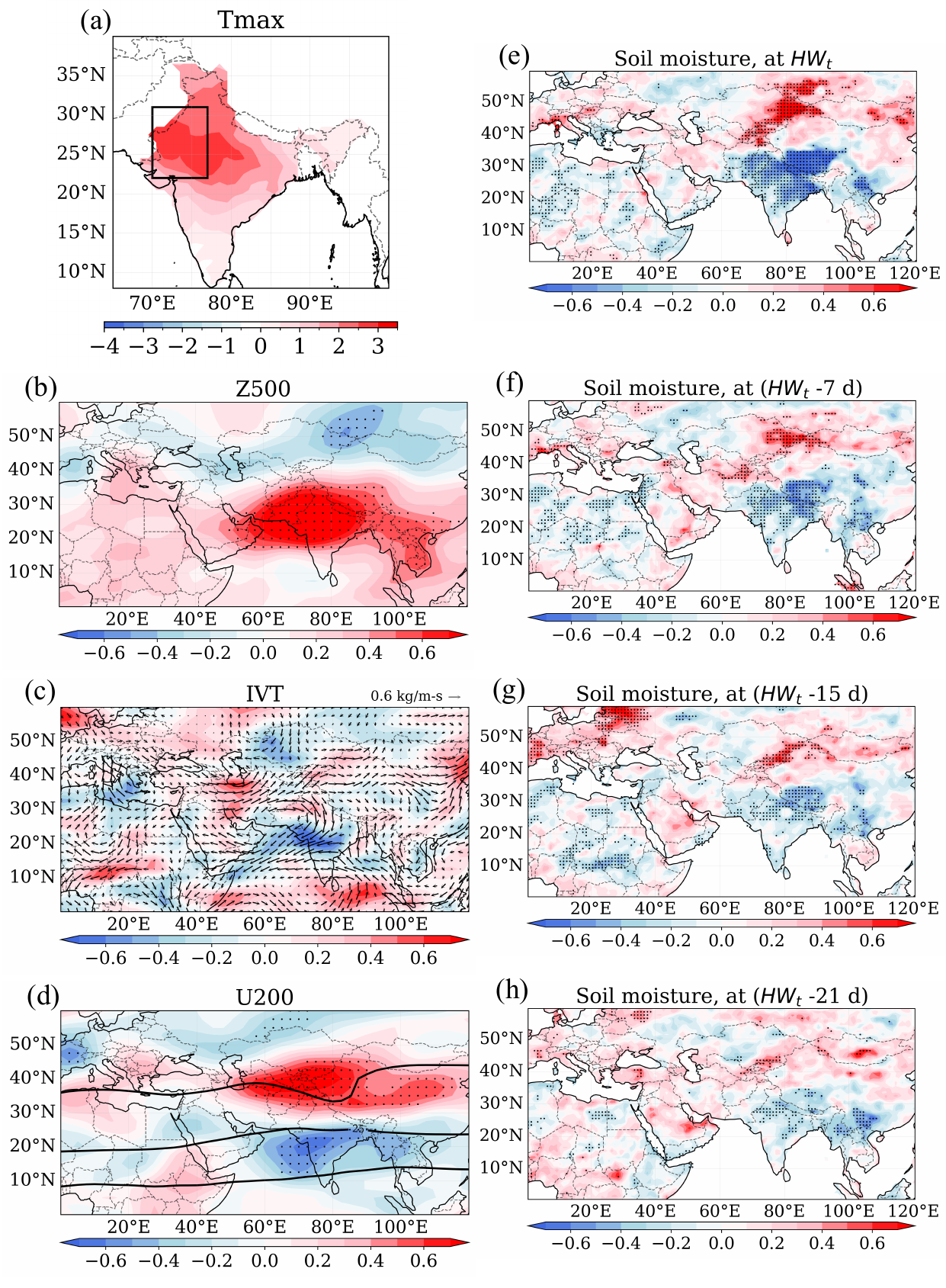}
\caption{Composites of different climate variables linked to HWs in the NCI.(a) maximum temperature [$^\circ C$], (b) 500 hPa geopotential height [$m$], (c) integrated moisture transport (IVT) [$kgm^{-1}s^{-1}$], and (d) zonal wind at 200 hPa ($U200$) [$m/s$] with black contours representing April-May climatology. Panels (e)-(h) depict soil moisture [$m^{3}/m^{3}$] anomalies occurring days to weeks before HWs, as indicated by the titles. Dotted areas highlight anomalies significant at the 95\% level. }
\label{fig:fig01}
\end{figure}

Our study begins by examining the atmospheric conditions linked to heatwaves in the NCI region (black rectangle in Figure ~\ref{fig:fig01}a), utilizing NCEP-NCAR 1 reanalysis data. Figure~\ref{fig:fig01} (b-g) shows the standardized composite anomalies of various atmospheric parameters such as Figure \ref{fig:fig01}a daily maximum temperature ($T_{max}$), Figure \ref{fig:fig01}b mid-tropospheric geopotential height ($Z500$), Figure \ref{fig:fig01}c vertically integrated moisture transport (IVT), Figure \ref{fig:fig01}d upper tropospheric zonal wind ($U200$) and Figure \ref{fig:fig01}e soil moisture (SM), all occurring at the time of HW onset. The figure shows that, coinciding with the HWs, a high-pressure system develops over India (Figure \ref{fig:fig01}b). Consistent with ~\cite{ratnam2016anatomy,rohini2016variability}, this pressure anomaly is associated with anticyclonic downward motion of air masses, producing dry air and reducing moisture flow over India (Figure ~\ref{fig:fig01}c). Figure~\ref{fig:fig01}d  suggests a simultaneous northward shift in zonal wind at 200 hPa ($U200$) with respect to climatology (solid black contours denote April--May $U200$ climatology). 

As low soil moisture can influence the atmospheric circulation patterns through land-atmospheric feedback~\citep{fischer2007soil,koster2016impacts}, in Figures~\ref{fig:fig01} (f - h) we plot the evolution of the composites of the anomalous SM prior to the HW events at NCI.
Generally, a strong relationship between SM and surface temperature can be observed over Indian subcontinent~\citep{berg2015interannual,miralles2012soil,joy2022influence,koster2006glace}. Reduced soil moisture decreases latent heat flux while increasing surface warming; which in turn amplifies sensible heat flux through diabatic heating, leading to elevated air temperatures~\citep{alexander2011extreme,rothlisberger2023quantifying,fischer2007contribution,rohini2016variability,seneviratne2010investigating,perkins2015review}.  Low SM conditions are evident as early as 3 weeks prior to the occurrence of a HW. As we approach the HW onset ($HW_t$), dry moisture conditions intensify. Such persistent dry soil moisture (SM) conditions can also be linked to low late winter--spring precipitation (February to April) as shown in Figure~\ref{fig:sup_preci}. 

\begin{figure}[!ht]
\centering
\includegraphics[width=\linewidth]{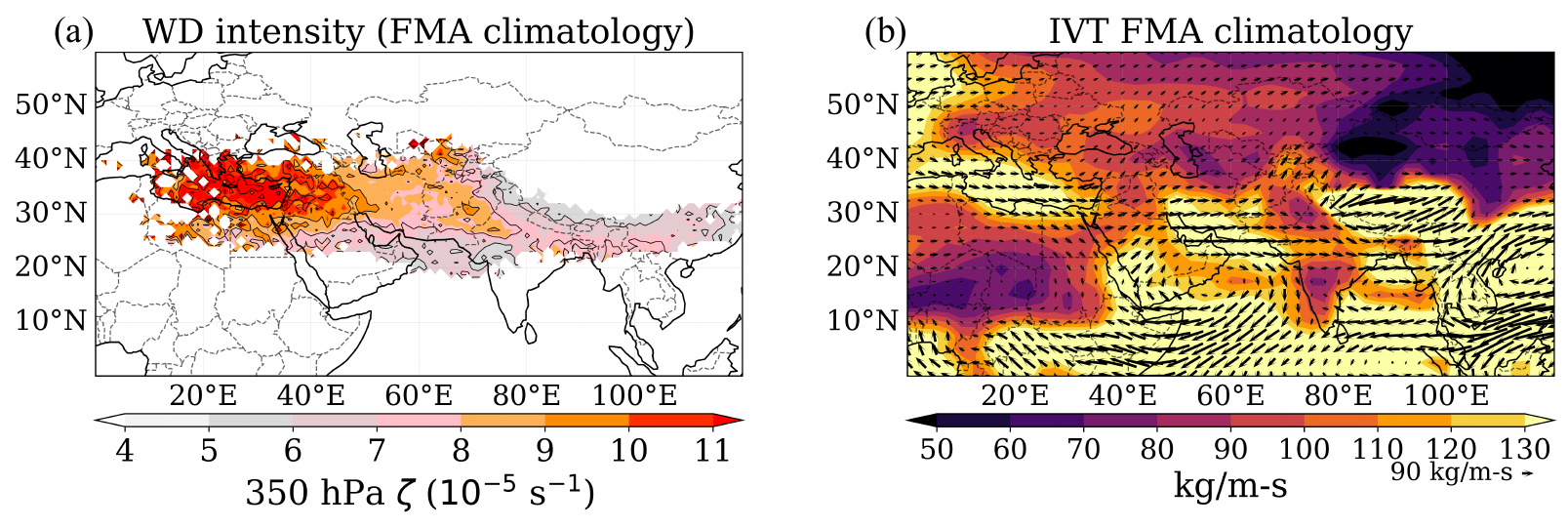}
\caption{(a) February--April (FMA) climatology of the western disturbances (WDs) (350 hPa relative vorticity) derived from catalogue provided by~\citet{hunt2023linking}. (b) Climatology of the vertically integrated moisture flux during same period.}
\label{fig:fig02}
\end{figure}

To gain further insight into the factors contributing to low SM leading to subsequent HW development in NCI, we investigate the moisture dynamics in the context of large-scale circulations during the period of February to April (FMA). The moisture transport during winter--spring is generally associated with the intensity and position of the `western disturbances' (WDs)~\citep{hunt2018extreme,hunt2023linking,yadav2012characteristic}.
The climatology of Western Disturbances (WDs) and vertically integrated water vapour transport (IVT) during February, March, and April (FMA) is illustrated in Figure~\ref{fig:fig02}(a, b). The climatology of WD intensity shows high values over the Mediterranean Sea, Iraq, Iran, Pakistan, and the northwest of India (Figure~\ref{fig:fig02}a). These intense WDs transport moisture from the Mediterranean, Red Sea, Persian Gulf, and northern Arabian Sea to northwest of India and the Himalayan region, continuing further east as shown in Figure~\ref{fig:fig02}b~\citep{baudouin2021synoptic}. The variability in moisture transport over northern India is often associated with fluctuations in the intensity and frequency of WDs~\citep{hunt2024western}.

Previous studies have reported that the position and intensity of the subtropical westerly jet (SWJ) influence WDs activity~\citep{filippi2014multidecadal,dimri2015western,hunt2023linking,hunt2018evolution}. The position of the SWJ influences the baroclinic interaction between WDs and the low-level moisture flow from the Arabian Sea~\citep{hunt2024western}. \citet{hunt2023linking} found that a northward shift in the SWJ position causes a poleward shift in WDs activity. Such a poleward shift in the SWJ has been previously linked using paleoclimatic records with hot and dry spring in the north-western Himalaya region~\citep{thapa2020poleward}. The variability of the SWJ is closely related to the climate conditions over its North Atlantic entry point to Eurasia~\citep{hunt2023linking}.

Here we use EOF analysis of $U200$ anomaly during FMA in the North Atlantic to investigate the relationship between the jet position and the dynamics of moisture flow which are associated with anomalously low soil moisture conditions over northern India.
\begin{figure}[!ht]
\centering
\includegraphics[width=\linewidth]{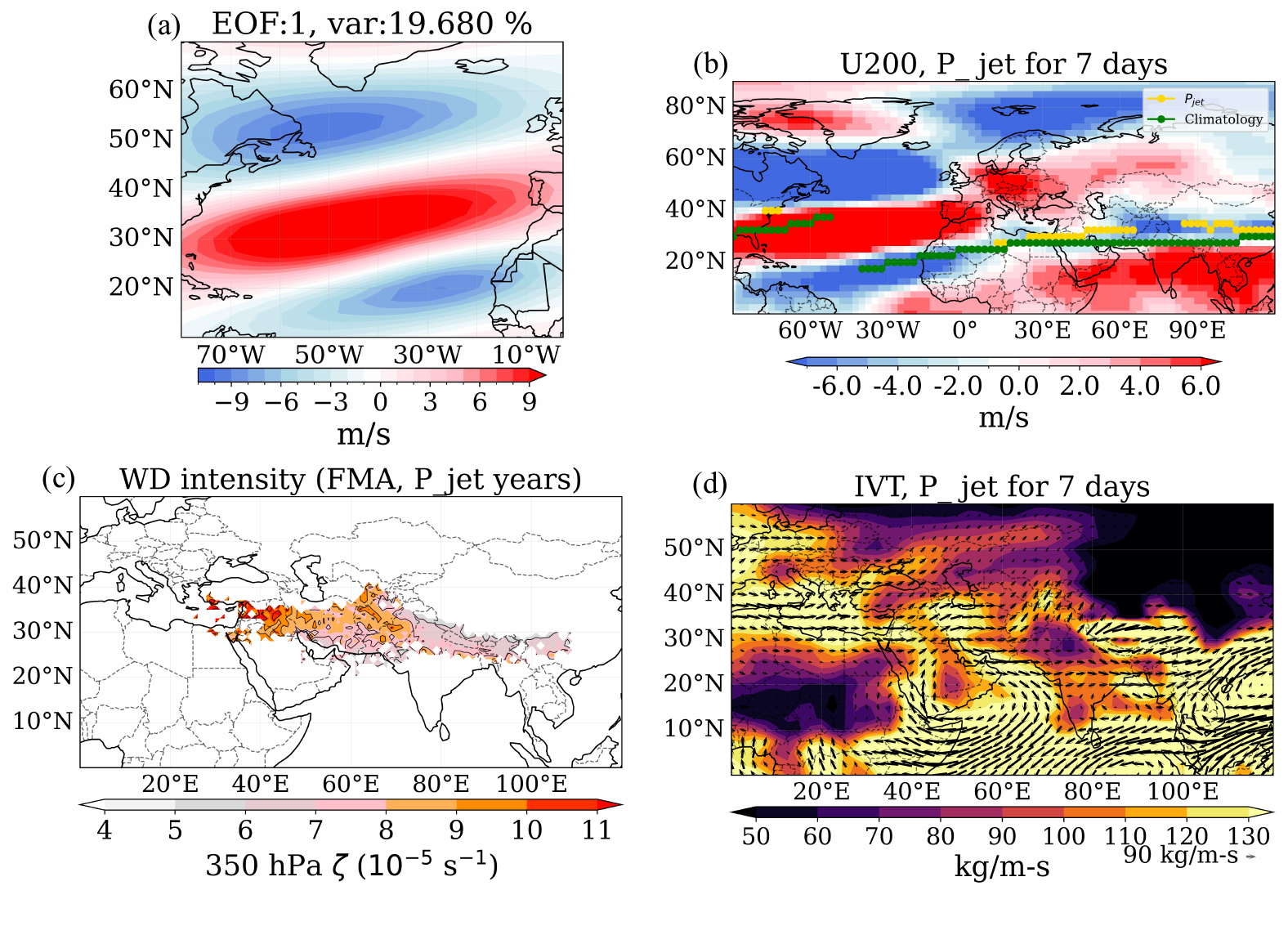}
\caption{(a) EOF 1 of the $U200$ anomaly during February-March-April (FMA). (b) $U200$ anomaly associated with the $90^{th}$ percentile of PC1 persists for 7 days. Green dots indicate the climatology of the SWJ, and yellow dots denote the SWJ associated with $P_{jet}$ indices. The position of the SWJ is computed by taking a meridional local maximum in 200 hPa wind speed ($U$) where $U$ exceeds 30 m/s~\citep{hunt2023linking}. (c) Mean WDs during FMA for $P_{jet}$ years. (d) Moisture flux (IVT) associated with $P_{jet}$ indices.}
\label{fig:fig03}
\end{figure}

The first EOF (EOF1) of the $U200$ anomaly during FMA in the North Atlantic region, shown in Figure \ref{fig:fig03}a, explains $19\%$ of the variance. The principal component time series (PC1) associated with EOF1 positively correlates with the jet position and waviness. High positive values of the PC1 are found to be related to the northward shift of the jet~\citep{hunt2023linking}. Then, we use the PC1 time series and construct $P_{jet}$ index which computes the time indices when PC1 is persistently high for 7 days above a threshold (here it is 90\% of the PC1). The $U200$ anomaly associated with the $P_{jet}$ indices indicates a northward movement in $U200$ (Figure \ref{fig:fig03}b). In Figure \ref{fig:fig03}b, green dots denote the climatology of the SWJ and yellow dots indicate the SWJ associated with $P_{jet}$ indices.
 
The moisture flow (IVT) associated with the $P_{jet}$ indices shows an increased influx of moisture from the Mediterranean Sea, Black Sea and Caspian Sea to the northwest part of Iran, parts of Afghanistan, Tajikistan and the western Himalayan region (Figure \ref{fig:fig03}d). In contrast, low moisture flow is observed over northwestern India (Figure \ref{fig:fig03}d), compared to the climatology (Figure \ref{fig:fig02}b). We see a reduced moisture flow in the southern parts of Iran,  which may be linked to a decrease WD activity in this region (Figure \ref{fig:fig03}c). Furthermore, the influx of moisture from the north of the Arabian Sea to southern Pakistan and northwest India is reduced during the presence of persistent jet conditions. The $P_{jet}$ indices can therefore identify the times when the connection between upper-level synoptic-scale disturbances and low-level moisture transport from the Arabian subtropical gyre weakens, providing valuable insights into moisture flow dynamics in the region. 

\begin{figure}[!ht]
\centering
\includegraphics[width=\linewidth]{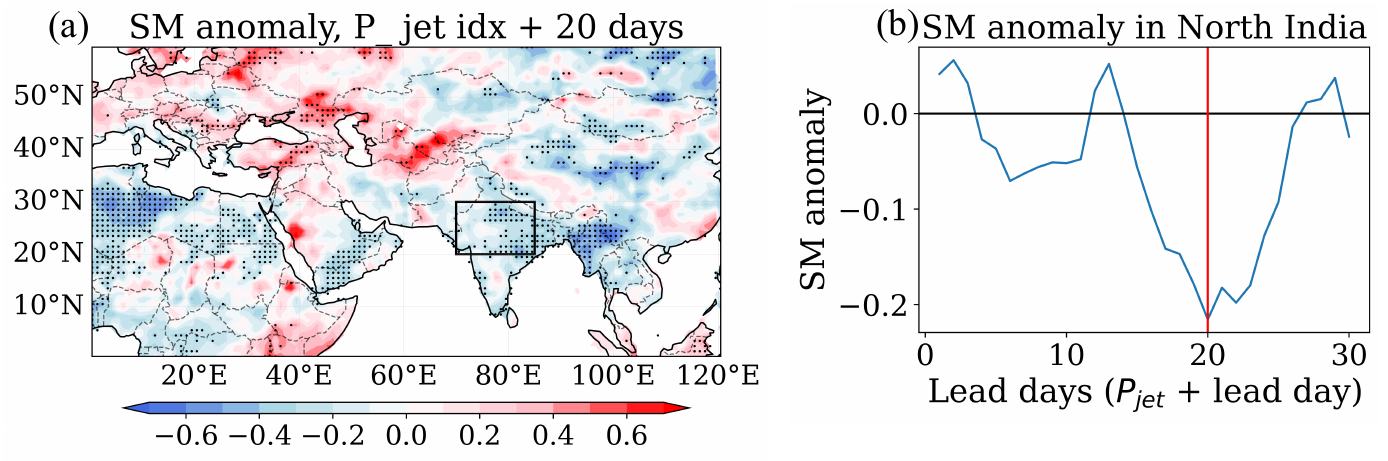}
\caption{(a) Soil moisture anomaly [$m^{3}/m^{3}$] 20 days (vertical red line in (b)) after the occurrence of $P_{jet}$. (b) SM anomaly [$m^{3}/m^{3}$] over the North India (black box in (a)) for different lead times, red vertical line denotes the 20 day after the occurrence of $P_{jet}$}
\label{fig:fig04}
\end{figure}

Using the $P_{jet}$ indices we can study the impact of the change in moisture dynamics associated with persistent jet conditions on soil moisture variability over India. Soil moisture anomalies over northern India exhibit a lagged response to $U200$ in the Atlantic region (Figure \ref{fig:sup_sm_u200}). Specifically, positive $U200$ anomalies in the North Atlantic during February--March is positively correlated with negative soil moisture conditions over northern India during April--May. We also find that the soil moisture negative response to the $P_{jet}$ is lowest at 20 days after the occurrence of $P_{jet}$ (vertical red line in Figure \ref{fig:fig04}b). Figure \ref{fig:fig04}a shows the composites of anomalous soil moisture fields associated with 20 days after the occurrence of $P_{jet}$. A significantly negative soil moisture anomaly can be observed over northern India.

Identification of the occurrence of the persistent jet using EOF-based analysis of zonal wind over the North Atlantic provides crucial insights into moisture flow dynamics and subsequent soil moisture variability. $P_{jet}$ indices enable us to distinguish the favourable atmospheric circulation patterns that influence local climate conditions typical for HW occurrence over India (Table \ref{table1}). Approximately 84.62\% of the $P_{jet}$ events occurred during a heat wave (HW) year. Furthermore, 53.85\% of the $P_{jet}$ events occurred during a HW in the northwestern part. When considering the total number of recorded HW years, the percentage of $P_{jet}$ events during a HW in the northwestern part is approximately 33.33\% (see Tables 4.2 and 4.3 in \citet{rajeevan2023heatwavescoldwaves}). \citet{rousi2022accelerated} associated an increase in the frequency and persistence of double jet stream states over Eurasia with occurrence of western Europe HWs. However, it should be noted that the identification of $P_{jet}$ indices, based on a 7-day persistent high value of PC1 and its response to low soil moisture in India within the selected time window of 20 days, is empirically derived. 

\begin{table}[ht]
\caption{List of $P_{jet}$ indices year, frequency, HWs}
\centering
\label{table1}
\resizebox{\textwidth}{!}{
\begin{tabular}{|l| c| c| c|}
\hline
\textbf{$P_{jet}$ occurrence} & \textbf{Frequency} & \textbf{Heat wave onset} & \textbf{Areas affected} \\
\hline
1980-02-01, 1980-02-08 & 2 & 18 April  & Northwest India \\
\hline
1981-03-04, 1981-03-11, 1981-03-18 & 3  & 15 Jun & Northwest India\\
\hline
1985-02-05 & 1 & 23 April, 1st May & East coast of India\\
\hline
1987-02-13 & 1 & 7 April & East coast of India \\
\hline
1988-04-11 & 1 & 10 April, 8 May, 26 May & Northwest India \\
\hline
1995-02-01 & 1 & 30 May & Northwest India \\
\hline
1997-04-16 & 1 & 29 May & East coast of India \\
\hline
2005-03-09 & 1 & 19 June & Northwest India \\
\hline
2006-03-22 & 1 & -- & None reported \\
\hline
2007-02-05, 2007-02-16 & 2 & 15 May & East coast of India \\
\hline
2010-02-03, 2010-02-10, 2010-02-17, 2010-03-03 & 4 & 8 April, 19 May, 19 June; 1 April, 9 April & Northwest India, East coast\\
\hline
2013-03-02, 2013-03-23 & 2 & 18 May & Northwest India\\
\hline
2018-03-02, 2018-03-09 & 2 & -- & None reported \\
\hline
\end{tabular}
}
\end{table}

In summary, the observational study using reanalysis data highlights the role of local soil moisture in the development of heat wave conditions in the NCI region. It suggests that these extreme heat events during April-May are driven not only by local weather patterns on a weekly scale but also by a complex interplay of large-scale factors, such as the persistence of jet stream states, which operate over monthly timescales. Therefore, it is crucial to investigate whether these factors have a causal impact on the occurrence of extreme temperatures in northern India. In the following section, we utilize an adjoint-based approach and perturbation experiments to explore the sensitivity of extreme temperatures to these factors.

\section{Revealing causal drivers of NCI Heatwave: Model-based study}\label{model_study}
\subsection{Adjoint sensitivity analysis}

\begin{figure}[!ht]
\centering
\includegraphics[width=\linewidth]{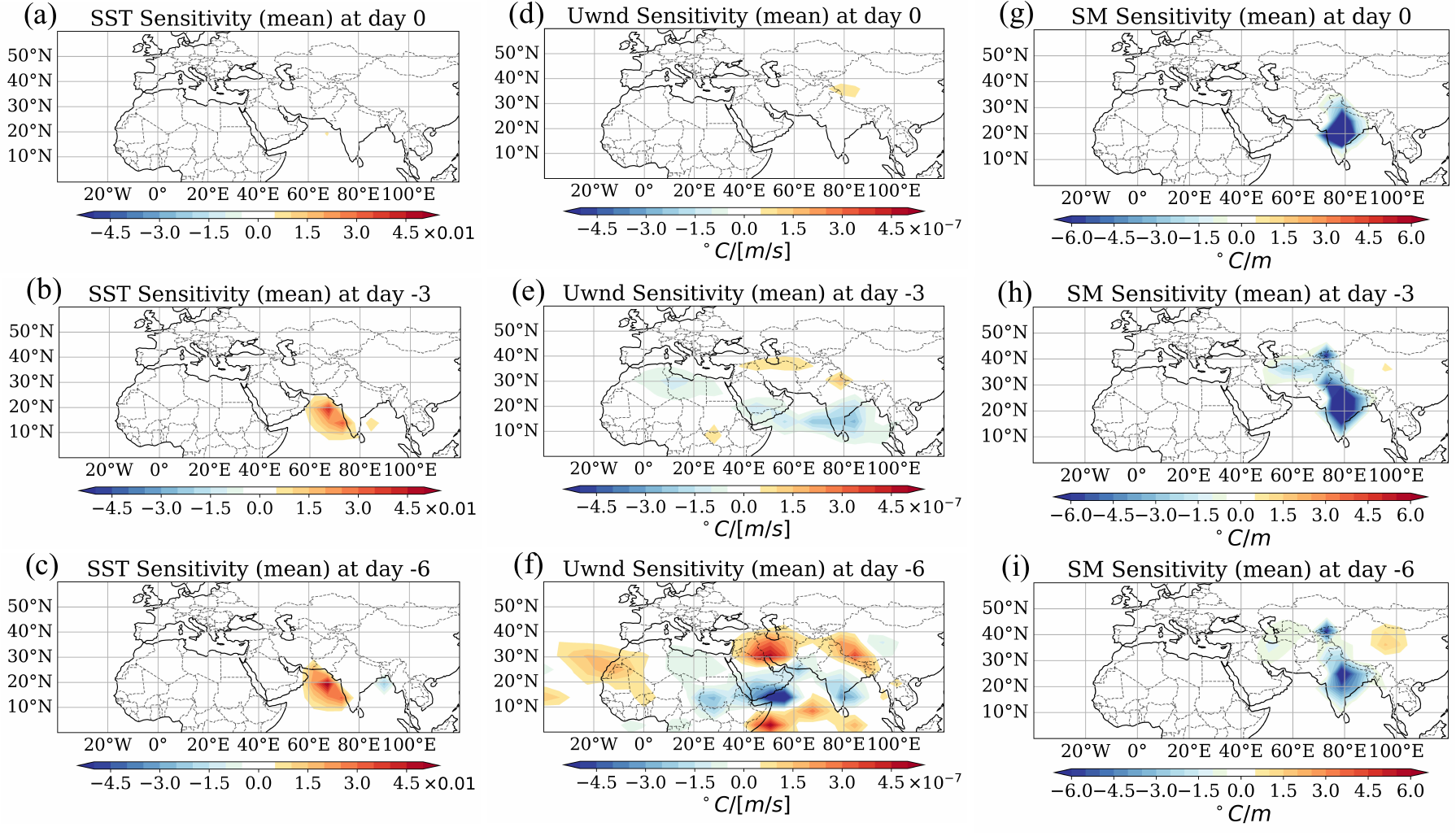}
\caption{Adjoint sensitivities of the cost function eq.~(\ref{eq1}) plotted backward in time from day 0 in the top row to -6 days in the bottom row. The left column (a - c) shows sensitivities with respect to for SST. The middle column (d - f) shows the sensitivities with respect to the upper tropospheric zonal wind (model level 2 analogous to 200-250 hPa). The right  column (g - i) shows the sensitivities with respect to soil moisture.}
\label{fig:ad02}
\end{figure}

Adjoint sensitivities can identify local and remote influences of model parameters on the cost function specified as the area-averaged $T2m$ over northern India (Figure~\ref{fig:ad01} and eq.\ref{eq1}). Figure \ref{fig:ad02} displays the ensemble-mean adjoint sensitivity of $T2m$ averaged over the masked region with respect to a few model parameters, going backward in time from $t_{0}$ to $t_{{0} }- 6$ days, where $t_0$ is the HW onset date. 

The sensitivity to ocean surface temperature (SST) can be identified over the specified time frame to originate mainly from the Arabian Sea (Figures~\ref{fig:ad02}(a - c)). The figure reveals that the $T2m$ sensitivity to $SST$ is very small at day 0, but grows as we go back in time and it can be anticipated that it amplifies further with increasing lead times. Positive $SST$ sensitivity originates in the north east of the Arabian Sea expanding up to the North of the Indian Ocean and south west of the Bay of Bengal for day -3 to -6. Positive sensitivity implies that the temperature increment in the masked region is associated with positive fluctuation in the $SST$ over this region. Previous studies have also reported a significant positive influence of $SST$ over the Arabian Sea and the western Indian Ocean on the temperature in northwestern India~\citep{rohini2016variability,dubey2021understanding}.     

With respect to upper tropospheric zonal wind, we observe small positive sensitivity to wind anomalies over the western Tibetan Plateau, some part of the north India, northern Pakistan on the day of HW onset (day 0) as shown in Figure~\ref{fig:ad02}d. By day -3 (Figure~\ref{fig:ad02}e), the positive sensitivity over northern India expands, while the negative sensitivity begins to extend from central India towards Oman, Yemen and northern Africa. Upon integrating further backward (day -6), the negative sensitivity expands further west, and positive sensitivity is observed over the Atlantic towards the Atlantic entry point of the African jet, with additional low negative sensitivity observed over the west coast of India.
Meanwhile, positive sensitivity stretches further eastward and westward from the Himalayan region. Notably, positive sensitivity is seen over Iran, Mediterranean regions. An increasing positive sensitivity is also observed over the western Indian Ocean region (Figure~\ref{fig:ad02}f). These sensitivity patterns (Figures \ref{fig:ad02}d-f) show a resemblance with the $U200$ composite shown in Figure~\ref{fig:fig01}d. The positive sensitivity over northern India indicates that zonal wind over this region is positively associated with the HW development in the north-central India. This can be linked with the shift in the zonal wind, as observed in Figure~\ref{fig:fig01}d and~\ref{fig:fig02}b. Despite the limitation in extending the analysis further backwards in time due to instability in the adjoint model, the upper tropospheric zonal wind sensitivity clearly indicates a robust link with remote large-scale circulation originating from the North Atlantic, thereby supporting our findings on the role of wind patterns associated with HWs as explained above.

A strong negative sensitivity of $T2m$ to the soil moisture (SM) can be seen over central India on the day of HW onset (Figure~\ref{fig:ad02}g). This negative sensitivity pattern indicates that dry local soil moisture strongly influences the HW development in this region which maintains our inference from the SM in the previous section (Figure~\ref{fig:fig01}h). As we integrate backward in time, negative sensitivity weakens and also spreads to the north of India, northern Pakistan, and southwestern Tibetan Plateau (Figures~\ref{fig:ad02}h,i). 
It is evident from both the observation and the adjoint-based sensitivity study, persistent low soil moisture is an important pre-requisite for the development of HWs. However, the biggest impact of SM on $T2m$ appears to come 1 day prior to the onset of heat wave; prior to this soil moisture could be more important to reshape the atmospheric flow via land-atmospheric feedback processes. It must be noted that the relative importance of the individual sensitivities (Figure~\ref{fig:ad02}) of the cost function can only be inferred qualitatively in this case as it requires the knowledge of expected anomalies to quantify their impact.   
We will analyse the individual contributions to a HW in the following step. 

From the adjoint sensitivity analysis above, we see that soil moisture has a localized dominant and robust affect on $T2m$ few days before the HW event. It is therefore of interest to quantify the relative contribution of SM to the temperature anomaly during a HW event. The response function of the linear adjoint model to SM can be evaluated by computing the dot product between the anomaly (here, the deseasonalized and detrended SM anomaly from the reanalysis data) and the sensitivity field, e.g., SM sensitivity ($\frac{\delta T2m}{\delta SM}$)~\citep{kohl2019seasonal}. The response function is computed as follows:
\begin{equation*}
\text{Response} \% = 100 \times \frac{\left( \frac{\delta T_{2m}}{\delta SM} \bigg|_{t'} \right) \cdot \ \text{SM}_{\text{anomaly}}(HW_t-t')}{\max\limits_{HW_t \leq t \leq HW_t+10} T_{2m \ \text{anomaly}}(t)\bigg|_{\text{NCI}}}
\end{equation*}
where  $t'$ denotes the day of minimum sensitivity for soil moisture,  $SM_{\text{anomaly}}$ is the observational soil moisture anomaly,  $HW_t$  is the onset of the HWs provided by the IMD (Section \ref{data_method}), and  $T_{2m, \, \text{anomaly}}$ is the temperature anomaly at NCI (Figure \ref{fig:fig01}a). We consider the maximum temperature anomaly between the onset and 10 days after the onset of the HW for the response function.

\begin{figure}[!ht]
\centering
\includegraphics[width=\linewidth]{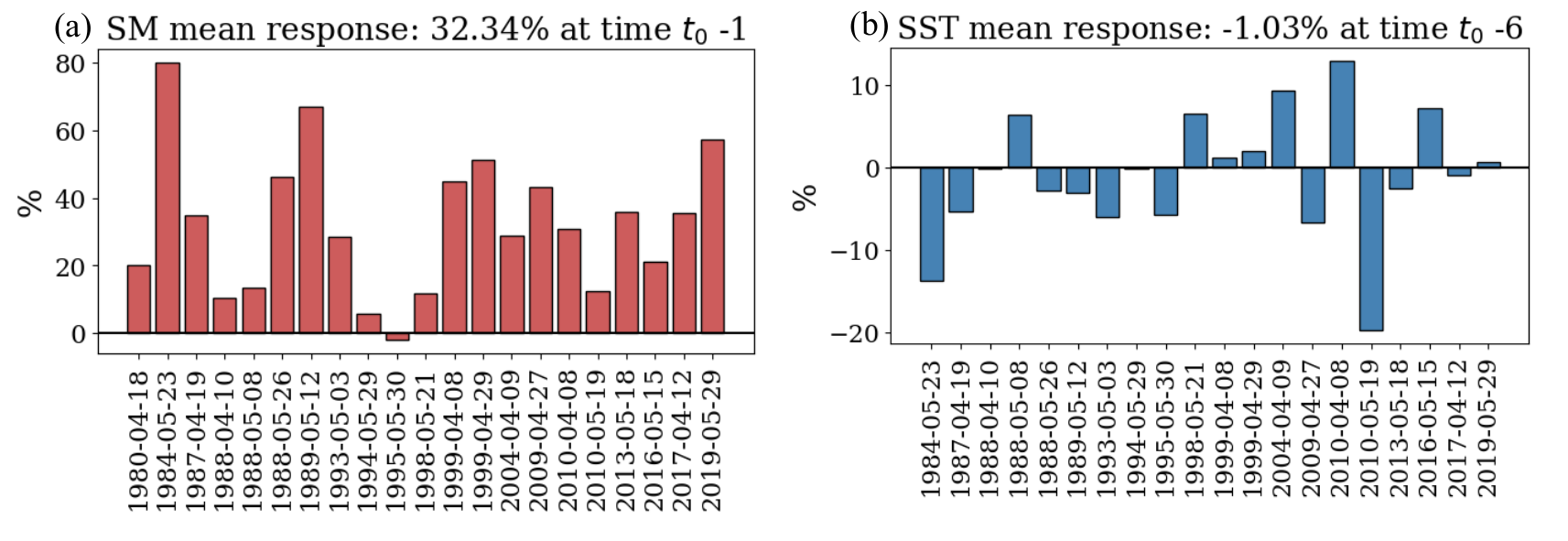}
\caption{(a) Soil moisture contribution to the temperature anomaly for different HW (-5 to 0 days, where Day 0 is the HW onset) events is shown}
\label{fig:SM_Response}
\end{figure}

For the soil moisture sensitivity field, we find that the minimum sensitivity occurs at $t'$ = -1 . Figure \ref{fig:SM_Response}(a) shows that the mean contribution of soil moisture to the temperature anomaly during all heatwaves in NCI is approximately 32\%. Additionally, we examine the SST response function and find that the maximum sensitivity for SST occurs at  $t'= -6$ , as shown in Figure \ref{fig:SM_Response}(b). We also investigate the contribution of upper-tropospheric zonal wind, but the responses are very small.


\subsection{Forward perturbation runs}
As the adjoint cannot be run backwards for longer than 1 week because of the fast growing nonlinear modes, it is difficult to infer the sensitivities of soil moisture over NCI stemming from regions further remote. However, we can demonstrate these dependencies in forward perturbation runs. 

In Section \ref{results}, we observed that the variability of the SWJ during FMA is associated with hot and dry conditions over India during April--May. The adjoint sensitivity analysis also revealed a strong connection to the SWJ (Figures \ref{fig:ad02}e,f). Several studies have associated changes in the sea surface temperature (SST) of the tropical Atlantic to a strengthening of the subtropical jet to the north~\citep{brayshaw2009tropical}. In agreement with these studies, we also observed a positive SST anomaly in the North Atlantic region during the occurrence of the $P_{jet}$ (Figure \ref{fig:sup_sst_anom}). It is therefore intriguing to investigate whether the North Atlantic SST plays indeed a role in moderating the large-scale circulations and whether these changes lead to dry soil conditions thereby exacerbating high temperatures over India. 

\begin{figure}[!ht]
\centering
\includegraphics[width=0.5\linewidth]{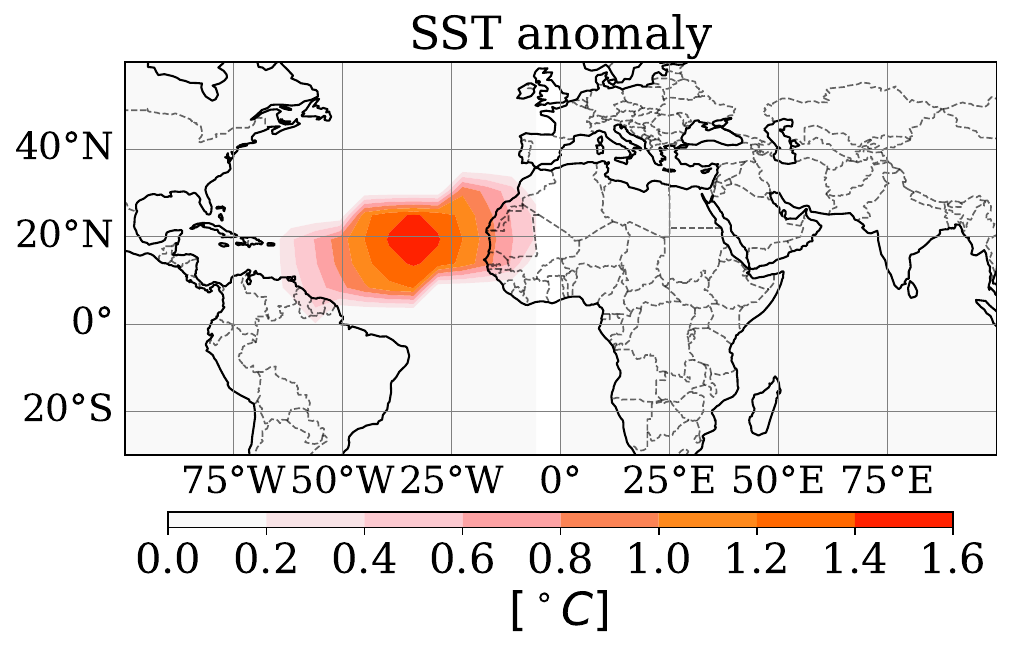}
\caption{SST perturbation mask for forward runs. }
\label{fig:sst_mask}
\end{figure}

\begin{figure}[!ht]
\centering
\includegraphics[width=\linewidth]{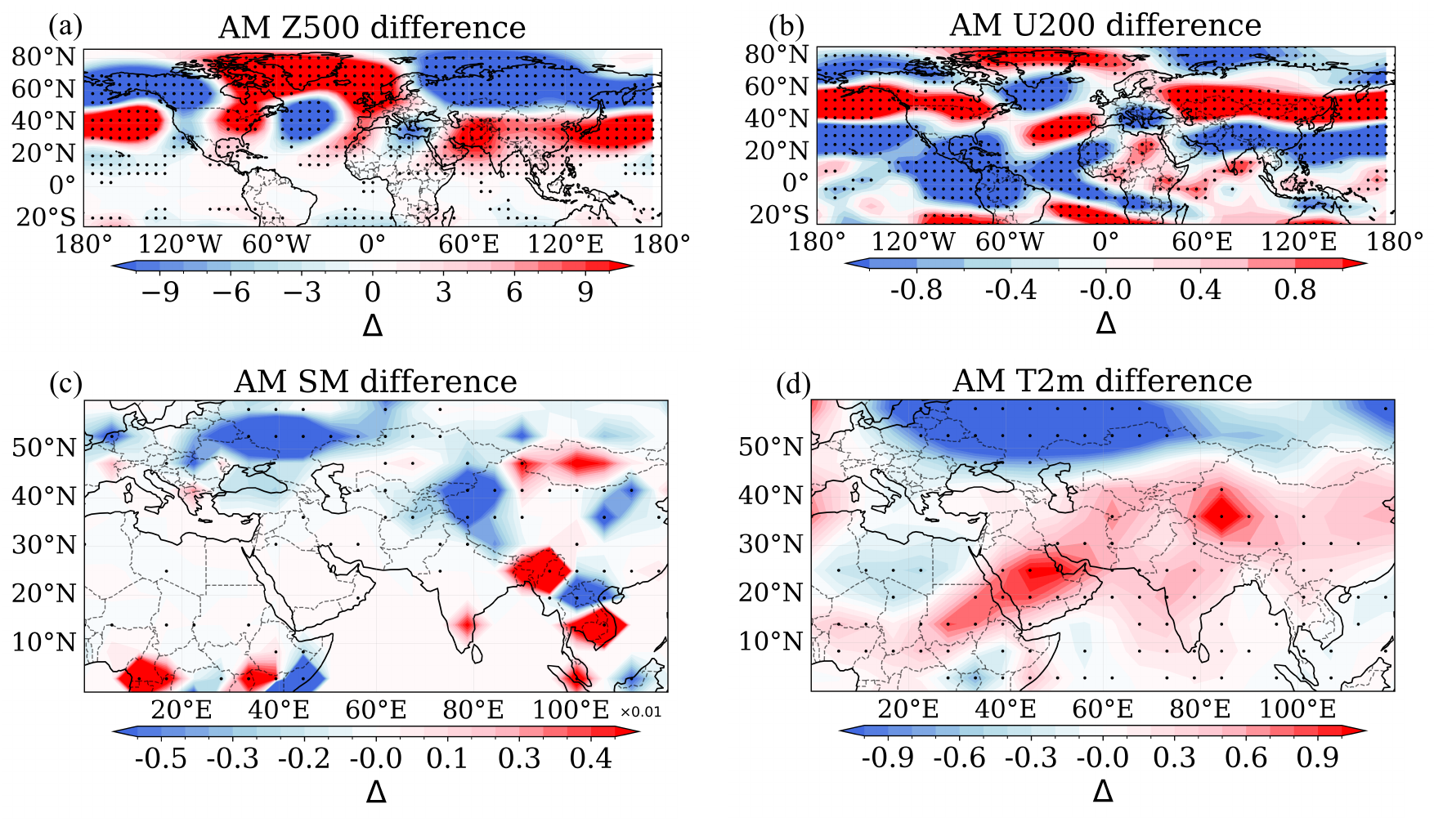}
\caption{Ensemble mean difference for different variables during April and May (AM) for forward run experiments in which the North Atlantic SST was perturbed (as shown in Figure~\ref{fig:sst_mask}) in the beginning of February: (a) $Z500$ [$m$], (b) $U200$ [$m/s$], (c) soil moisture [$m$], (d) 2m air temperature [$^\circ C$]. Black dots denote 95\% significance.}
\label{fig:fwd01}
\end{figure}

To investigate this question, we added a positive anomaly to the initial state to the North Atlantic SST in PlaSim's. We introduce this anomaly by creating a weighted mask (Figure \ref{fig:sst_mask}), which is then added to the initial SST values in the North Atlantic Ocean separately for February and March. Thereafter, we run the model forward up to the month of May. PlaSim has a slab ocean model of mixed layer depth of 50 meter. We repeat this experiment for 20 different starting years with similar perturbation adding to their initial values (20 ensemble members). Since we are investigating the heatwaves occurring in April-May, we calculate the mean difference of various variables between the perturbed and unperturbed SST conditions during this period. The motivation behind perturbing the SST of the North Atlantic during this time is drawn from our results presented in  Section 3, where we were able to link the occurrence of persistent jet in Feb-Mar to the occurrence of HW events in Apr-May over India (Table \ref{table1}). 

\begin{figure}[!ht]
\centering
\includegraphics[width=\linewidth]{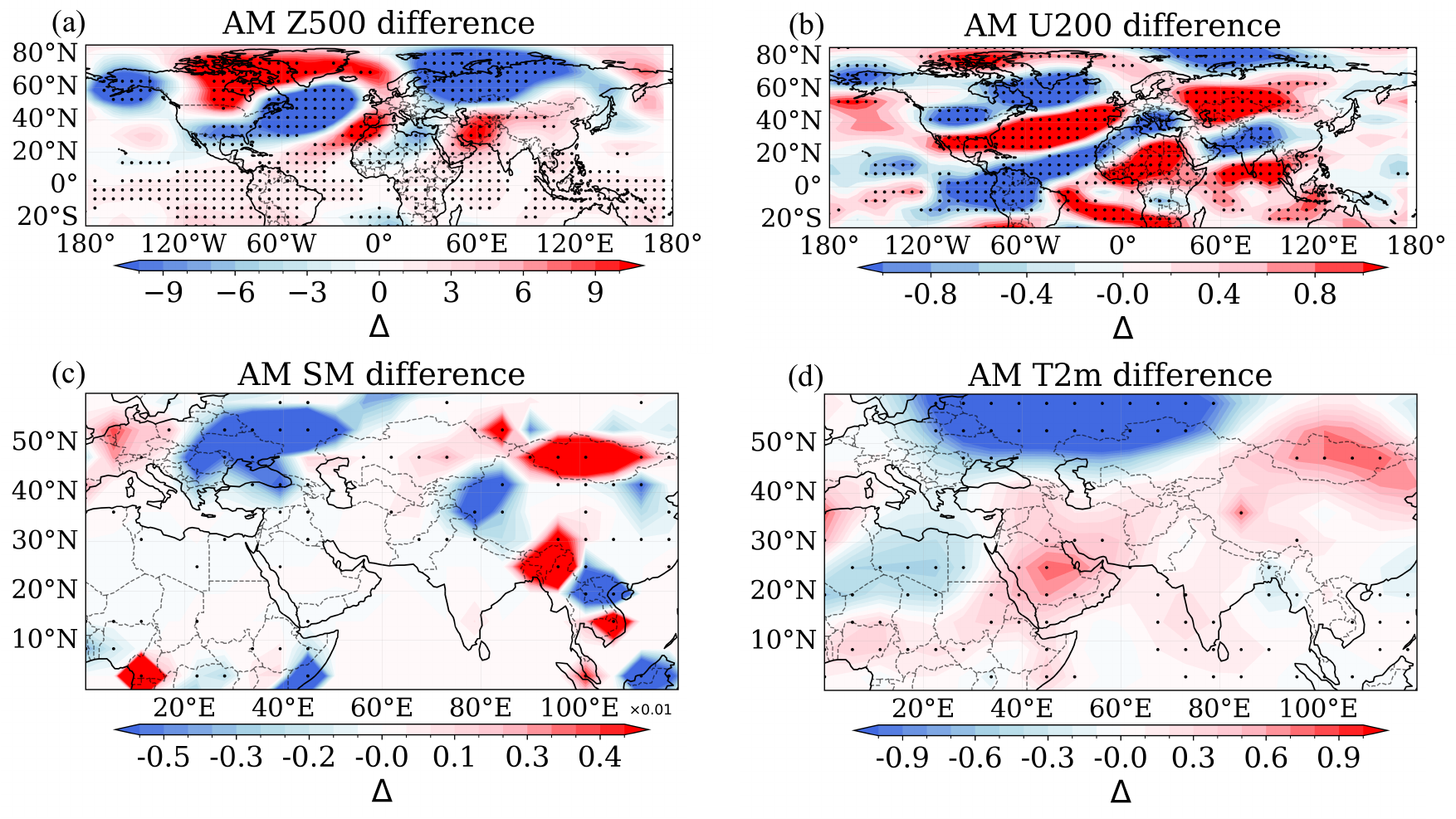}
\caption{Ensemble mean difference for different variables during April and May (AM) for forward run experiments in which the North Atlantic SST was perturbed (as shown in Figure~\ref{fig:sst_mask}) in the beginning of March: (a) $Z500$ [$m$], (b) $U200$ [$m/s$], (c) soil moisture [$m$], (d) 2m air temperature [$^\circ C$]. Black dots denote 95\% significance.}
\label{fig:fwd02}
\end{figure}

The ensemble mean difference between normal atmospheric state and the conditions on perturbing north Atlantic SST (Feb/Mar) during April-May (AM) is shown in Figures \ref{fig:fwd01} and \ref{fig:fwd02}. These figures demonstrate change in  geopotential height at 500 hPa (Z500), upper tropospheric zonal wind at 200 hPa (U200), 2m air temperature (T2m), and soil moisture (SM). Upon perturbing SST during both February (Figure \ref{fig:fwd01}a) and March (Figure \ref{fig:fwd02}a) we observe a high pressure anomaly over the southeast of Iran, Pakistan, northern India, northwest of Europe, and low pressure over the Eurasian region. Z500 patterns exhibit a quasi-stationary Rossby wave pattern. These findings corroborate with~\citep{ratnam2016anatomy} who also reported a causal relationship between high pressure system over northwestern Europe and India and a quasi-stationary Rossby wave pattern during HWs in NCI region. 

The highest zonal wind anomaly can be seen over the Eurasian region (mainly over Kazakhstan, Kyrgyzstan, Uzbekistan) for both the February (Figure \ref{fig:fwd01}b) and the March (Figure \ref{fig:fwd02}b) perturbation. We also observe significant negative anomaly over central India. These patterns bear close resemblance with the northward shift U200 pattern  associated with HWs in NCI shown in Figure \ref{fig:fig01}d. We also observe a significant negative soil moisture anomaly stretching over Tibetan Plateau, Himalayan region (Figures \ref{fig:fwd01}c and \ref{fig:fwd02}c). This pattern has similarity with the SM conditions prior and during HW events (Figures \ref{fig:fig01}e,f,g,h) and those $\approx$ 3 weeks after $P_{jet}$ (Figures \ref{fig:fig04}a). On the other hand, a positive soil moisture anomaly pattern is seen over northern Pakistan extending north eastwards to Kazakhstan, Russia and Mongolia, which may be related to changed moisture dynamics associated with the northward shift of SWJ in previous months as reported in Section 3 (Figure \ref{fig:fig03}d). This dipole-like pattern in SM anomaly is similar to that observed from reanalysis data weeks before and during HW events (Figures \ref{fig:fig01}e,f,g and h). The disparity between the intensity of negative SM anomaly over Indo-Gangetic plain and Himalayan belt, and the rest of India, in particular, a very weak signal of drier than normal SM conditions on perturbation of Atlantic SST might be partly associated with the fact that here we compute the mean SM conditions over a span of two months and partly due to model limitations. The latter includes coarse grid resolution, and generally weak moisture dynamics and precipitation biases.

Finally, Figures \ref{fig:fwd01}d and \ref{fig:fwd02}d show an overall significant positive anomaly of T2m over India in both cases of perturbation. The higher temperature anomaly observed in the northeastern parts of India, compared to the rest of the country, can be attributed to factors similar to those previously discussed in relation to the negative soil moisture (SM) anomaly.
It is important to note that despite the limitations of our simulations, such as the intermediate complexity of the model and the coarse resolution, our findings are consistent with those from our observational study detailed in Section 3. Our forward experiments demonstrate the significant influence of North Atlantic Ocean sea surface temperatures (SST) on large-scale atmospheric circulation over India. Specifically, our findings reveal that the dynamical changes in the atmosphere, driven by SST variations during February and March, can lead to increased temperatures over India in April-May through modulation of the upper tropospheric winds.

\section{Conclusion}
In this paper we provide a new working hypothesis for the development and causes of heatwaves over north-central India during April--May based on a combined analysis of atmospheric reanalysis fields and adjoint model-based approaches. Through our reanalysis study, we identified potential mechanisms leading to heatwaves over northern India. In combination with our novel adjoint sensitivity approach and perturbation runs, we identify the possible causal influence of local and remote factors causing extreme heat conditions in India.
Our results highlight a sequence of processes leading to heatwave development over this region begins to unfold months before the heatwaves actually occur. 

The proposed mechanism of heatwave development over India is summarized as follows:

\begin{enumerate}
\item Anomalously high sea surface temperatures occurring over the North Atlantic during February and March influence the variability of the subtropical westerly jet leading to teleconnections that can modulate weather conditions over the Indian subcontinent. 
\item Atmospheric dynamical changes in the North Atlantic lead to a northward shift in the position of the subtropical jet. Perturbation forward experiments show that anomalous SST within specific regions of the North Atlantic lead to changes in large-scale atmospheric patterns over monthly timescales, resulting in increased temperatures over northern India. The intensity and spatial extents of the SST anomalies applied in this study were empirically determined. 
\item This altered state of the subtropical jet can persist over several days which affects the area and intensity of the western disturbances embedded in it. 
\item Change in the upper tropospheric conditions reduces winter--spring moisture flow to the northern India. As western disturbances generally bring moisture in the Middle East and northwestern parts of India India during boreal winter--spring, the altered upper tropospheric conditions influence the moisture dynamics in this region. This includes reduced moisture flow to northern India from the north of the Arabian Sea during winter--spring. 
\item Reduced moisture flow affects local soil moisture variability creating conditions conducive for heatwave development. Adjoint sensitivity analysis indicated that dry soil moisture prior to the onset of heatwaves strongly influences heatwave development, even though the backward integration was limited up to 6 days due to nonlinear instability of the adjoint model. Moreover, significant reduction in soil moisture was observed from reanalysis study weeks before the onset of April--May heatwaves in India. 
\item Persistent low moisture before a heatwave can act as a preconditioning factor, establishing ideal conditions for heatwaves development. Consequently, extreme heat conditions occur during April-May over north-central India.
\end{enumerate} 

Identifying the preconditioning factors that lead to heatwave development is crucial for making accurate sub-seasonal predictions. In this context, it would be highly valuable to assess and evaluate the effectiveness of these factors in predicting heatwaves using high-resolution Earth system models and machine learning techniques. Further research is needed to explore the effects of varying the intensities and spatial extents of the SST in the North Atlantic to fully understand their impact on heatwaves over India. A rigorous analysis is also required to fully uncover the relationship between 
$P_{jet}$ and soil moisture variability over India, which is beyond the scope of this study.

\clearpage
\section*{Acknowledgement}
 This research was supported, in part, through the Koselleck grant $\text{Earth}^{RA}$
from the Deutsche Forschungsgemeinschaft. Contribution to the Centrum für Erdsystemforschung und Nachhaltigkeit
(CEN) of Universität Hamburg.
%
%
\section*{Data}
We utilize daily maximum temperature data over India from 1980 to 2022, as provided by the India Meteorological Department~\citep{srivastava2009development}. For atmospheric variables, we use daily data from the NCEP-NCAR Reanalysis 1 for the same time period~\citep{kalnay2018ncep}. Daily surface soil moisture data is sourced from GLEAM v3.8a~\citep{martens2017gleam}. Precipitation data is obtained from the GPCC monthly precipitation product~\citep{schneider2016gpcc}\footnote{\url{https://opendata.dwd.de/climate_environment/GPCC/html/fulldata-monthly_v2022_doi_download.html}}. Additionally, we use the NOAA Optimum Interpolation (OI) sea surface temperature (SST) V2 high-resolution daily data from 1982 to 2022~\citep{huang2021improvements}, regridded to the PlaSim grid for SST response function analysis, and the NOAA Extended Reconstructed SST (V5) monthly data for composite analysis~\citep{huang2017noaa}.

\section*{Appendix}
\renewcommand{\thefigure}{A\arabic{figure}}
\setcounter{figure}{0}
We identify the heatwave years and compute the composites of standardize precipitation anomaly for April-May and also the months before.
\begin{figure}[!ht]
\centering
\includegraphics[width=\linewidth]{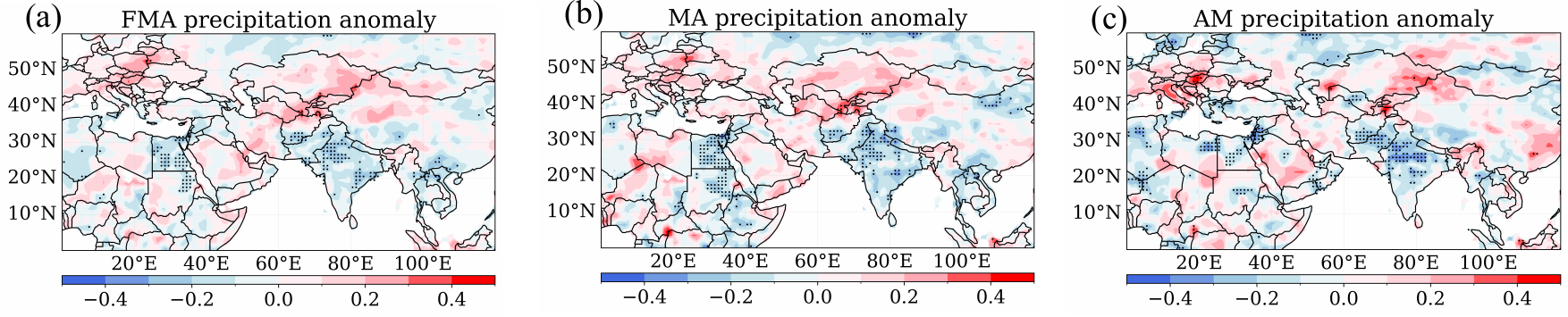}
\caption{GPCC monthly precipitation [$mm/month$] anomaly on HW years, (a) February-April, (b) March-April, (c) April-May }
\label{fig:sup_preci}
\end{figure}

\begin{figure}[!ht]
\centering
\includegraphics[width=\linewidth]{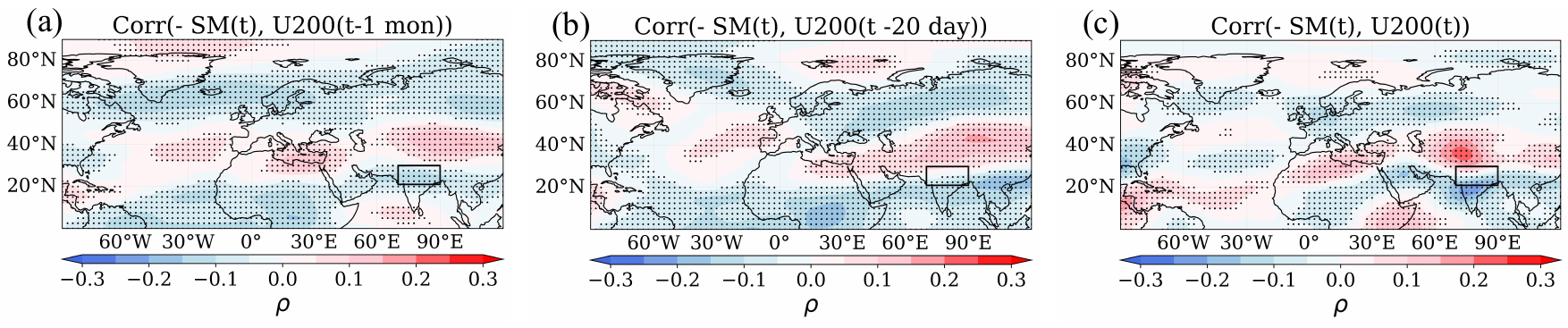}
\caption{April-May Correlation between Soil moisture anomaly over north India and $U200$ with lags, $\rho$ denotes the correlation strength, black dots imply $95\%$ significance.}
\label{fig:sup_sm_u200}
\end{figure}

We compute the SST anomaly using the long term climatology between 1991-2020 provided by \url{https://downloads.psl.noaa.gov/Datasets/noaa.ersst.v5/}
\begin{figure}
\centering
\includegraphics[width=\linewidth]{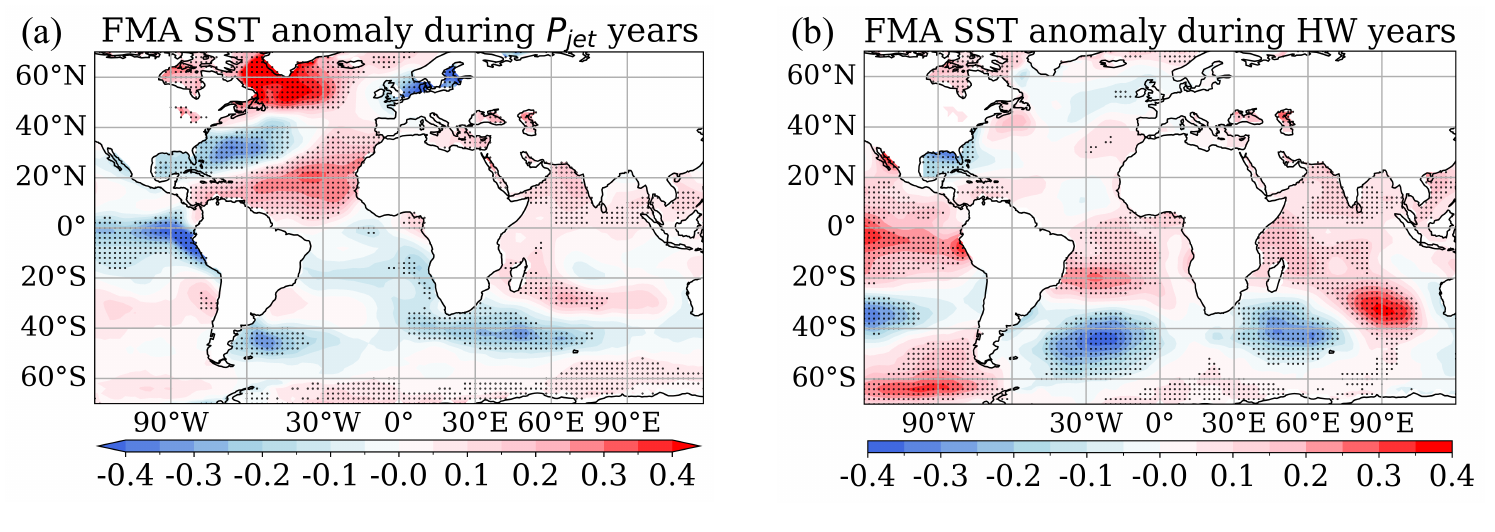}
\caption{Monthly sea surface temperature anomaly [$^\circ C$] during February-March-April (FMA) for (a) $P_{jet}$ years, (b) HW years. Black dots imply $95\%$ significance.}
\label{fig:sup_sst_anom}
\end{figure}

\clearpage
\bibliographystyle{apalike}
\bibliography{reference}
\end{document}